\documentstyle[12pt]{article}
\newcommand{\bey}[1]{\begin{eqnarray} \label{#1}}
\newcommand{\eey}{\end{eqnarray}}
\newcommand{\beq}[1]{\begin{equation} \label{#1}}
\newcommand{\eeq}{\end{equation}}
\newcommand{\degree}{$^\circ$}
\title{A Configurational Bias Monte Carlo Method
for Linear and Cyclic Peptides}

\author{Michael W.\ Deem$^*$ and Joel S.\ Bader\\
CuraGen Corporation\\
322 East Main Street\\
Branford, CT\ \ 06405}

\begin{document}
\maketitle
\renewcommand{\baselinestretch}{1.3}
\tiny
\normalsize
~\\
\\ Running Title: Biased Monte Carlo of Cyclic Peptides.\\
~\\
~\\
\noindent$^*$Author to whom correspondence should be addressed.
Present address: Lyman Laboratory of Physics,
Harvard University, Cambridge, MA\ \ 02138.
\newpage

\begin{center}Abstract\end{center}
In this manuscript, we describe a new configurational bias Monte Carlo
technique for the simulation of peptides.  We focus on the
biologically relevant cases of linear and cyclic peptides.  Our approach
leads to an efficient, Boltzmann-weighted sampling of the torsional degrees
of freedom in these biological molecules, a feat not possible with previous
Monte Carlo and molecular dynamics methods.
%
%
%We present a new configurational bias Monte Carlo method for the
%simulation of linear and cyclic peptides.  The method employs
%importance sampling to achieve efficient equilibration at the
%relevant body temperature.  We argue that
%standard Monte Carlo or molecular dynamics techniques would fail to
%equilibrate the important case of cyclic peptides.\\
\newpage

\section{Introduction}
\hbox{}
This paper presents a new Monte Carlo method that employs biased
trial moves to achieve an efficient sampling of the torsional degrees
of freedom for linear and cyclic peptides.

Peptides are small molecules, built from amino acids, that are of
fundamental importance in biological systems \cite{Alberts}.  They play
key roles in signal transduction between cells, regulation of cell growth and
differentiation, and protein localization on cell surfaces
\cite{Cohen}.  Peptides
are thought to regulate neurotransmission, from modulating
pain and thirst to affecting
memory and emotion \cite{Kandel,Pert}.
  They are
used as a chemical defense mechanism by some organisms.  The {\em conus} snails,
for example, produce a family of highly-constrained peptides that include
very powerful neurotoxins \cite{Olivera}.  Finally, peptides are used
within the biotechnology industry to identify
antagonists blocking various abnormal enzymatic actions or
ligand-receptor interactions \cite{Clackson}.  Cyclic or otherwise
constrained peptides are often preferred for this application, since such
molecules suffer less of a loss of configurational entropy upon binding
\cite{Alberg}.  A classic example is the use of the RGD peptide to block
the GPIIb/IIIa-fibronectin interaction, reducing blood platelet aggregation
\cite{Ruoslahti,ONeil}.

The properties of peptides are amenable to examination by computer
experiment.  An early study was of the alanine dipeptide, in
which the potential energy surface was deduced from {\em ab initio}
quantum mechanical calculations \cite{Cheam,Tobias}.
Larger peptides have been
examined by classical simulations.  Both molecular dynamics \cite{Roux}
and Monte
Carlo \cite{Nikiforovich} approaches have proven useful.  The effects of
the aqueous environment have been incorporated by simple dielectric
theory \cite{Schiffer,Smith,Gould,Daggett}
or by explicit inclusion of water molecules \cite{Yan}.

It has become clear, however, that the standard molecular dynamics
and Monte Carlo methods are not capable of sampling all
conformational degrees of freedom accessible at body temperature to
the larger peptides.  This problem is particularly evident for the
important case of constrained peptides.  Various solutions, such as
high-temperature molecular dynamics \cite{Bruccoleri,Tsujishita} or simplified force
fields \cite{Tsujishita,Brunne}, have been suggested, but these approaches
suffer from uncontrolled approximations.  A simulation method able to
sample the relevant conformational states of peptides, particularly
constrained ones,  or exposed loops of larger proteins
 would be of great value.  It would aid study
of these molecules in biological systems as well as facilitate
structural understanding of the
peptides and antibodies of interest to the biotechnology industry.

Recently, powerful Monte Carlo methods have been developed that have a
greatly enhanced sampling efficiency
\cite{Dodd,Frenkel,SmitV,dePablo,Smit,Maginn,Leontidis,Escobedo}.
  These methods have been applied
to chain molecules at low and high density \cite{SmitV,dePabloII} and even at phase
coexistence \cite{SmitIV,dePabloIII,SmitIII,SmitII}.
These methods all use importance sampling, or
biased moves, to efficiently explore the free energy landscape.

We here apply these concepts to peptide molecules.  Both linear and
constrained or cyclic peptides are treated by this method. 
In Sec.\ 2 we describe the Monte Carlo method in detail.
Appendices describe the rigid molecular fragments from which
peptides are constructed and provide technical details
of the method.
  In Sec.\ 3 we describe the application of this method to the
prototypical polyglycine peptides.  We discuss the results in Sec.\ 4. 
The superiority of this method over conventional molecular dynamics and
Monte Carlo is demonstrated.
Conclusions are presented in Sec.\ 5.

\section{Monte Carlo Method}

We make the simplifying assumption that the intramolecular potential energy
contains only torsional and non-bonded terms.  That is, bond lengths and
angles are fixed, and rotation is allowed only about sigma bonds.  At room- or
body-temperature, these are fairly good assumptions. They could
easily be relaxed, although sampling the increased degrees of freedom would
entail a computational expense.
Appendix A describes the rigid fragments that occur in peptides under these
assumptions.  A suitable form for the interatomic potential would be the
AMBER \cite{Weiner}, ECCEP
\cite{Nemethy}, or CHARMm \cite{Brooks} force field.  We pick the AMBER
potentials.  Water is treated in an implicit way, assuming the
dielectric constant for Coulomb interactions is given by
$\epsilon/\epsilon_0 = 4 r$, with $r$ given in {\AA}ngstroms. 
These assumptions allow the method to be presented without a
discussion of detailed force field issues.  The method is generically
applicable to better force fields and an explicit treatment of water.

A configurational bias Monte Carlo (CBMC) technique is used to
explore the conformations of the molecules.  We describe the
algorithm for both linear and cyclic peptides.  By cyclic, we mean
peptides constrained because of disulfide bonds between cystine
residues.

There are two types of atoms in a peptide, those in the side chains and
those in the backbone.  Consequently, there are two types of Monte Carlo
moves: type I moves change the positions of side chain atoms only, and type II
moves change the positions of backbone atoms, rigidly rotating the attached side chains.
The type I move
is an extension of the chain-molecule CBMC
\cite{SmitV,dePablo} to the structurally more complicated case of peptides.  The
type I move is applicable to side chains with a free end ({\em i.e.}\ all
naturally occurring amino acid side chains except for proline).  The
backbone to which the side chain is attached can be either linear or
cyclic.  In the cyclic case, the type I move
is also used to change the configuration of
the free ends of the main chain.

There are two kinds of type II moves for the backbone: type IIa moves for
linear peptides and type IIb moves for cyclic peptides.  The type IIa move
is essentially the same as a type I move.
The side-chain residues that are attached to the backbone are rigidly rotated
so as to remain properly bonded to the C$_\alpha$ atoms in their new
positions.
When the peptide is cyclic, we use a type IIb move to change the
configuration of part of the backbone loop, rigidly rotating any
side chains or free ends of the peptide that are attached to that part of
the backbone.  The backbone of a cyclic peptide includes the atoms along
the main chain as well as the C$_{\beta}$ and S atoms of the cystines
participating in the disulfide bond.  This move requires a concerted
rotation of the backbone torsional angles with a rigid rotation of the
attached side groups.  This concerted rotation of the torsional angles is
an extension of the concerted rotation scheme for alkanes
\cite{Dodd,Leontidis}.  

A type I move is initiated by identifying the side chain to be regrown.
Not all of the side chain need be regrown, and the first group to regrow is
chosen.  This feature is helpful for the amino acids with longer side
chains, such as lysine.  These choices are made randomly.  The $M$ rigid
units to be regrown are first removed and then
added one at a time, starting from the one closest
to the backbone.  For each addition, the following actions are carried out
(see Fig. 1):

1) $k$ values of the torsional angle $\phi_{ij},~ 1 \le j \le k$  connecting rigid unit
$i$ to unit $i-1$ are generated according to the internal potential,
\beq{5}
p_i^{int}(\phi_{ij}) \propto \exp[-\beta u_i^{int}(\phi_{ij})] \ .
\eeq
The function  $u_i^{int}(\phi_{ij})$ is the part of the internal
energy that couples unit $i$ to the rest of the molecule (but
excluding units $i+1$ to $M$).  The inverse temperature is
given by $\beta = 1/k_B T$.

2) One of these is picked with probability
\beq{6}
p_i^{ext}(\phi_{ij}) = \exp[-\beta u_i^{ext}(\phi_{ij})] / w^{ext}(i) \ ,
\eeq
where
\beq{7}
w^{ext}(i) = \sum_{j=1}^k  \exp[-\beta u_i^{ext}(\phi_{ij})]  \ .
\eeq
The function  $u_i^{ext}(\phi_{ij})$ is the part of the external
energy that couples unit $i$ to the rest of the molecule (but
excluding units $i+1$ to $M$).

3) Steps 1-2 are repeated until all M units have been added.

4) The Rosenbluth weight
\beq{8}
W^{(n)} = \prod_{i=1}^M w^{ext}(i)
\eeq
is calculated.  This attempted move is accepted with a probability
\beq{9}
acc(o \rightarrow n) = \min[1, W^{(n)}/W^{(o)}] \ .
\eeq
The quantity $W^{(o)}$ is the Rosenbluth weight for the reverse
move and is calculated as in steps 2-4, but with $k-1$ random
orientations and one orientation that is equal to the original
geometry for each rigid unit.

A type IIa move is very similar to a type I move.  In this case, the
direction of regrowth is chosen randomly.  Then the first backbone
unit to be regrown is chosen.  The $M$ rigid units to be regrown are
removed and added back sequentially, as in the type I move.  The
rigid units in this case are either A-units, B-units with the side
chain rigidly attached, C-units, or D-units (see appendix A).  An alternative
procedure would be to regrow the side chain units as well, but this
proved not to be efficient, due to frequent steric repulsions.  The
move is accepted with the probability given by Eq.\ (\ref{9}).

A type IIb move is initiated by identifying the 4 rigid units on the
backbone to be rotated.  This is done randomly.  The four rigid units
are labeled in an amine to carboxy terminal fashion.  The attached
side groups are rigidly rotated with the backbone units.

The rotation is carried out as follows (see Fig. 2):

1) The driver angle $\phi_0$ is changed by an amount $\delta \phi_0$, where
$-\Delta \phi < \delta \phi_0 < \Delta \phi$.
This is done $k'$ times with probabilities according to the internal
potential,
\beq{10}
p^{int}(\phi_{0j}) \propto \exp[-\beta u_0^{int}(\phi_{0j})] \ .
\eeq
The function  $u_0^{int}(\phi_{0j})$ is the internal energy associated
with this torsional angle.  Only those values of $\phi_0$ that lead
to valid solutions for the modified torsional angles are considered. 
In the general case there will be a distinct $\phi_1$ for each
solution arising from the new value of $\phi_0$.  Define $k^{(n)}$ to
be the number of $\phi_0$-$\phi_1$ pairs. If  $k^{(n)}=0$, the move
is rejected.

2)  A $\phi_0$-$\phi_1$ pair is
picked with probability
\beq{11}
p_0^{ext}(\phi_{0j}, \phi_{1j}) = \exp[-\beta u_0^{ext}(\phi_{0j},
\phi_{1j})] / W^{(n)} \ ,
\eeq
where
\beq{12}
W^{(n)} = \sum_{j=1}^{k^{(n)}}  \exp[-\beta u_0^{ext}(\phi_{0j},\phi_{1j})] \ .
\eeq
The function  $u_0^{ext}(\phi_{0j},\phi_{1j})$ is the part of the external
energy that couples this part of the backbone to the rest of the
molecule.  The value $J^{(n)}$ of the Jacobian is calculated for the new,
chosen configuration (as detailed in Appendix B).

3) The reverse move is considered.  That is, a rotation about the
new, chosen $\phi_0$-$\phi_1$ pair is considered.  $k'-1$ random values
$\delta \phi_0$ are chosen. The original value of $\phi_0$ is
assigned to the $k'$th value.  This move results in $k^{(o)}$
solutions for $\phi_1$.   $k^{(o)}$ is always greater than zero,
since the original configuration exists.  (Special care is taken to
ensure that the original configuration is found by the root finding
procedure.) The Rosenbluth weight is assigned to $W^{(o)}$.  The
value $J^{(o)}$ of the Jacobian is also calculated for the original
configuration.

This attempted move is accepted with a probability
\beq{113}
acc(o \rightarrow n) = \min[1, J^{(n)} W^{(n)}/ J^{(o)} W^{(o)}] \ .
\eeq

Splitting the energy into internal and external parts is rather
arbitrary.  There are some constraints imposed, however, by the
requirement that the normalization constants for Eqs.\ (\ref{5}) and
(\ref{10}) be independent of chain conformation \cite{Smit}.  We
assume for simplicity that $u_i^{int} = 0$.  One other natural
choice, however, would set the internal part equal to the torsional
terms in $H_{intra}$ and set the external part equal to the rest of
$H$.

For any Monte Carlo scheme to properly sample the Boltzmann
probability distribution, detailed balance must be satisfied.  Refs.\
\cite{Dodd} and \cite{Smit} prove that detailed balance is satisfied
for the above scheme.

\section{Application to Polyglycine}
In this section we present the result of applying this
configurational bias Monte Carlo method to two simple peptides,
polyglycine G$_6$ and constrained polyglycine CG$_6$C.

Figure 3 shows the energy of linear polyglycine as a function of
Monte Carlo steps.
  This run took roughly 3 hours on a
Silicon Graphics Indigo$^2$.
  In Fig.\ 4 we show the end-to-end probability
distribution for this system.
Gaining this degree of convergence took a one-day run.

Figure 5 illustrates the energy of the cyclic polyglycine as a function of
Monte Carlo steps.  This run took roughly 6 hours.
  Figure 6 provides a histogram of the number of
solutions found for each attempted concerted rotation.  In rare
cases the root finding procedure failed to find all the roots.  In the
construction of this plot, we rounded $n^{(n)}$ up when it was odd.
Figure 7 shows the
histogram for the C$_\beta$SSC$_\beta$
dihedral angle, with the statistics taken from a run six times as long as that
illustrated in Fig.\ 5.  To give a feel for the barrier to rotation
about this angle,
 we show in Fig.\ 8 the potential of mean
force.  This potential was
determined by umbrella sampling \cite{Chandler}. This curve took
two orders of magnitude
longer to determine than did the probability distribution in
Fig.\ 7.
 The potential of mean force is contrasted with the
energy associated purely with the ${\rm C_{\beta}SSC_{\beta}}$ torsional
terms.  Finally, Fig.\ 9  shows the result of classifying the configurations
produced by the method into distinct stable conformations.
Fuzzy clustering \cite{Gordon} was used to determine the dominant
conformations, with the result that there are only two or three distinct
conformations within this limited simulation run.  The simulation run
depicted in Figs.\ 5 and 9 took approximately 8 hours on a Silicon Graphics
Indigo$^2$.

\section{Discussion}

We see that with a very modest computational effort, we can achieve
equilibrated results for linear peptides.  With somewhat more effort,
we can achieve equilibration for cyclic peptides.

As expected, we find that the linear peptide G$_6$ is relatively
unstructured in solution.  There is a common crumpled
state, but there is also a significant population of the extended
state.  The constraint of the disulfide bond in CG$_6$C, in contrast,
forces that molecule to adopt a limited number of molecular
conformations.  For the fairly short runs illustrated in Figs.\ 5,6,7 and 9,
we find only three dominant conformations.  The first conformation
is associated with the C$_\beta$SSC$_\beta$
torsional angle of 290\degree,
whereas the other two are associated with angles of 88\degree\ and
98\degree.  The first of these conformations
is very tight, with 0.7 \AA\ fluctuations about the mean for all atoms
in the molecule.  The other two are somewhat looser, with roughly
1.2 \AA\ fluctuations.
We see from Fig.\ 9 that even in this short run the method
revisits previous conformations.  In the limit of a long simulation,
the time spent in each conformation would, of course, be proportional
to the exponential of the free energy of the conformation.

If  CG$_6$C were achiral, the potential of mean force in Fig.\ 8 would
be symmetric about 0\degree\ and 180\degree.  Since the C$_\alpha$ carbons
in the cystine residues are, in fact, chiral, the potential of mean force
is not required to be symmetric.  The asymmetry seen in Fig.\ 8 results from
the mean, chiral force of the rest of the molecule on the
C$_\beta$SSC$_\beta$ torsion.  In fact, the AMBER forcefield takes this
chirality into account by reducing the symmetry of the C$_\beta$ carbon in 
cysteine.  We have used this geometry \cite{InsightII}.  The barrier at 0\degree\ 
is due to a high steric repulsion between the hydrogens on the
C$_\beta$ carbons adjacent to the disulfide bond.  This barrier is
substantially higher than the barrier at 180\degree.

From Fig.\ 8, we see that there is a very significant free energy
barrier to rotation about the C$_\beta$SSC$_\beta$ torsional angle.
This figure was not constructed from a standard simulation run, but by the
specialized procedure of umbrella sampling.
It is clear
from Fig.\ 7, however, that the present method is able to overcome
this barrier and to properly sample the relevant conformations even in
a relatively short simulation.  Any method such
as molecular dynamics or standard Monte Carlo that makes only small,
local changes to the configuration would never cross this barrier in
a simulation of reasonable length.  
High temperature dynamics can allow
systems to cross high barriers, but can not perform the requisite Boltzmann
sampling to predict the physiologically relevant conformations.
Only a biased method that makes
fairly large geometrical changes is capable of dealing with such
barriers in an automatic way, without resort to special techniques
such as umbrella sampling.  Furthermore, the ability to perform umbrella
sampling has as a prerequisite the detailed knowledge of the important
conformations and the paths between them.  In our specific case, we find
our method to be two orders of magnitude more efficient than
umbrella sampling.

\section{Conclusion}

We have presented a Monte Carlo method capable of sampling the
relevant room- or body-temperature configurations of linear and cyclic
peptides.  This method allows the study of peptides important in 
biological and technological settings.
Our sampling of the disulfide dihedral angle in a prototypical
cyclic peptide indicates
that the method can explore widely separated regions of
conformation space according to the proper Boltzmann distribution, even if
the barriers between the regions are quite large.  Previous
simulation methods  either fail to sample the proper thermal distribution
or are vastly more computationally intensive and
require detailed knowledge of the thermally accessible regions.
The method can be extended to allow incorporation of
explicit water molecules.  The method can be extended to force
fields with flexible bonds and angles.  These extensions are subjects for
future work.

\section*{Acknowledgements}
We thank Berend Smit and Charlene X.\ L.\ Liang for helpful discussions
about the Monte Carlo method and Len Bogarad, Michael McKenna,
Jonathan Rothberg, and Gregory Went for helpful conversations
about the biological applications.  This work was supported by
the NCI/NIH under grant
%1 R43 
\#CA62752-01
and by the NIST ATP program under
grant number \#70NANB5H1066.
Many of the calculations described
herein were performed on an Indigo-R8000 on loan from SGI and on
a HP-735/125 on loan from Hewlett Packard.

\section*{Appendix A: Rigid Units}

As described, the algorithm assumes that bond lengths and angles are
fixed.  The only degrees of freedom, therefore, are torsional
angles.  Due to the extremely high force constant for rotation about
a $\pi$ bond, even some torsional angles are fixed as well.  An entire
collection of atoms that is rigid is called a rigid unit.  Such a
unit has an incoming bond as well as several possible outgoing bonds.
There are four backbone rigid units.  Unit A is the starting
NH$_3^+$ group.  Unit D is the terminal COO$^-$ group.  Unit B is the
C$_\alpha$H group.  Unit C is the CONH amide bond group.

The residues are connected to the backbone by outgoing bonds from the
B units.  Table 1 lists the decomposition of the amino acid side
chains into rigid units.  Typical rigid units are the CH$_2$,
CN$_3$, CO$_2$, and aromatic ring groups, which have
substantial $\pi$ bonding character.

Proline is a special case, technically an imino acid.  The special
nature is due to the cyclic bonding of the residue to the backbone. 
The rigid units in this amino acid are the CH$_n$, CO, and N groups. 
Only {\em trans} isomers are allowed for the proline
amide bond.  Proline is treated in an approximate way:  the 
C$_\alpha$-C$_\delta$ fragment is kept rigid, the
C$_\delta$-N bond is broken, and the C$_\alpha$-N torsional barrier
is increased.  This approximation ignores the small fluctuations in the
configuration of the proline side-chain loop.

\begin{table}
\caption
{
The rigid units in peptide side groups.
}
\begin{center}
\begin{tabular}{c c}
Side Group & Rigid Units\\
\hline
Glycine  &H\\
Alanine  &CH$_3$\\
Arginine  &CH$_2$, CH$_2$, CH$_2$, CN$_3$H$_5$$^+$\\
Aspartate &CH$_2$, CO$_2$$^-$\\
Asparagine & CH$_2$, CONH$_2$\\
Cyst(e)ine & CH$_2$, S(H)\\
Glutamate  &CH$_2$, CH$_2$, CO$_2$$^-$\\
Glutamine  &CH$_2$, CH$_2$, CONH$_2$\\
Histidine  &CH$_2$, C$_3$N$_2$H$_3$\\
Isoleucine  &CH, CH$_2$, CH$_3$, CH$_3$\\
Leucine  &CH$_2$, CH, CH$_3$, CH$_3$\\
Lysine  &CH$_2$, CH$_2$, CH$_2$, CH$_2$, NH$_3$$^+$\\
Methionine  &CH$_2$, CH$_2$, S, CH$_3$\\
Phenylalanine & CH$_2$, C$_6$H$_5$\\
Proline & Backbone Groups: C$_\alpha$HCH$_2$CH$_2$CH$_2$, N, CO\\
Serine  &CH$_2$, OH\\
Threonine & CH, CH$_3$, OH\\
Tryptophan &CH2, C$_8$NH$_6$\\
Valine  &CH, CH$_3$, CH$_3$\\
Tyrosine & CH$_2$, C$_6$H$_4$, OH\\
\end{tabular}
\end{center}
\end{table}

\section*{Appendix B: Concerted Rotation}

Since the molecules under consideration can be cyclic, a Monte Carlo
move that preserves this constraint is required.  The ``concerted
rotation'' scheme used for alkanes \cite{Dodd} can be
extended to allow rotation of the torsional angles in cyclic peptides.
This appendix describes this extension.  The reader is
referred to Ref.\ \cite{Dodd} for a fuller discussion of
the original, restricted method.
The method presented here allows for a fairly general molecular geometry.
In particular, the method naturally accommodates the constraint of a
planar amide bond.

To formulate the method, we consider rotating about seven torsional 
angles, which will move the root positions of four rigid units,
rotate up to three additional ones, and leave the rest of the
peptide fixed.  We define the root position of a rigid unit to be the
C$_\alpha$ position for a B unit, the C position for a C unit, the C
position for a CH$_2$ unit, and the S position for the S unit in
cystine.  If unit 5 is a C unit,  however,
${\bf r}_5$ is defined to be the N position of that unit.
  For each unit we define $\theta_i$ to be the angle between
the incoming and outgoing bonds.  Thus, $\theta_i = 0$ for a C unit,
and $\theta_i \approx 70.5^\circ $ for all others.  Figure 1
illustrates the geometry under consideration.

The method leaves the positions ${{\bf r}_i}$ of units $i\le 0$ or $i
\ge 5$ fixed.  The torsion $\phi_0$ is changed by an amount $\delta
\phi_0$.  The values of $\phi_i, 1 \le i \le 6$, are then determined
so that only the positions ${\bf r}_i$ of units $1 \le i \le 4$ are
changed.

The method required several definitions to present the solution
for the new torsional angles.  Vectors are defined which are
the difference in position between unit $i$ and unit $i-1$, as
seen in the coordinate system of unit $i$:
\beq{A2.1}
{\bf l}_i = {\bf r}_i^{(i)} - {\bf r}_{i-1}^{(i)} \ .
\eeq
The coordinate system of i is such that the incoming bond is along
the $\hat {\bf x}$ direction.  Thus ${\bf l}_i = l_i \hat {\bf x}$
if atom ${\bf r}_i$ and ${\bf r}_{i-1}$ are directly bonded and
has x- and
y-components otherwise.  We now define a rotation matrix that
transforms from the coordinate system of unit $i+1$ to unit $i$
\beq{A2.2}
{\bf T}_i = 
\left( \begin{array}{ccc}
\cos \theta_i & \sin \theta_i & 0 \nonumber \\[.2in]
\sin \theta_i \cos \phi_i & -\cos \theta_i \cos \phi_i & \sin \phi_i
\\[.2in]
\sin \theta_i \sin \phi_i & -\cos \theta_i \sin \phi_i & -\cos \phi_i
\end{array}
\right) \ .
\eeq

The positions of the units in the frame of unit $1$ are, thus, 
given by
\bey{A2.2a}
{\bf r}_1^{(1)} &=& {\bf l}_1 \nonumber \\
{\bf r}_2^{(1)} &=& {\bf l}_1 + {\bf T}_1 {\bf l}_2 \nonumber \\
{\bf r}_3^{(1)} &=& {\bf l}_1 + {\bf T}_1 ({\bf l}_2
   + {\bf T}_2 {\bf l}_3)  \nonumber \\
{\bf r}_4^{(1)} &=& {\bf l}_1 + {\bf T}_1 ({\bf l}_2
   + {\bf T}_2 ({\bf l}_3 + {\bf T}_3 {\bf l}_4))  \ .
\eey

We further define the matrix that converts from the frame of
reference of unit $1$ to the laboratory reference frame
\beq{A2.3}
{\bf T}_1^{lab} = 
 [\cos \psi {\bf I} + {\bf n} {\bf n}^\top
(1- \cos \psi) + {\bf M} \sin \psi] {\bf A} \ ,
\eeq
where
\beq{A2.4}
{\bf M} =
\left( \begin{array}{ccc}
 0 & -n_z & n_y \nonumber \\[.2in]
n_z & 0 & -n_x \nonumber \\[.2in]
-n_y & n_x & 0 \nonumber\\
\end{array}
\right) \ ,
\eeq
and
\bey{A2.4aa}
{\bf n} &=& \frac{\hat {\bf x} \times {\bf r}}
    {\vert \hat {\bf x} \times {\bf r}\vert } \nonumber \\
\cos \psi &=& \frac {{\bf r} \cdot \hat {\bf x}}
  {\vert {\bf r} \vert } \nonumber \\
\sin \psi &=& \frac {\vert {\bf r} \times \hat {\bf x} \vert}
  {\vert {\bf r}\vert  } \ ,
\eey 
where $\hat {\bf x}$ is a laboratory unit vector along the x direction, and
$\bf r$ is the axis of the bond coming into unit $1$.  The matrix
${\bf A}$ is a rotation about $\hat {\bf x}$ and is defined so that ${\bf
A}  {\bf l}_1 = \Delta {\bf r}$:
\beq{A2.4a}
{\bf A} =
\left( \begin{array}{ccc}
 1 & 0 & 0  \\[.2in]
0 & c & -s  \\[.2in]
0 & s & c
\end{array} \right)
\eeq
where
\bey{A2.4b}
  c &=& ({l_1}_y \Delta r_y + {l_1}_z \Delta r_z)/(\Delta r_y^2 + \Delta r_z^2)
\nonumber \\
 s &=& (-{l_1}_z \Delta r_y + {l_1}_y \Delta r_z)/(\Delta r_y^2 + \Delta r_z^2)
 \ .
\eey
Here 
$\Delta {\bf r} = {\bf A} [{\bf T}_1^{lab}]^{-1} ({\bf r}_1 - {\bf r}_0)$ if
unit 0 is a C unit; otherwise $\Delta {\bf r} = {\bf l}_1$.

The method proceeds by solving for $\phi_i, 2 \le i \le 6$, analytically
in terms of $\phi_1$.  Then a nonlinear equation is solved 
numerically to determine which
values of $\phi_1$, if any, are possible for the chosen value of
$\phi_0$.

We will work in the coordinate system of unit $1$, after it has been
rotated by the chosen $\phi_0$.  We define
\beq{A2.5}
{\bf t} = {\bf r}_5^{(1)} - {\bf l}_1 = [{\bf T}_1^{lab}]^{-1}
 ({\bf r}_5 - {\bf r}_0)  - {\bf l}_1 \ .
\eeq
If $\theta_3 \ne 0$ and  $\theta_5 \ne 0$, the square distance
between unit $3$ and unit $5$ is known and equal to 
\beq{A2.6}
q_1^2 = ({l_4}_x \cos \theta_4 - {l_4}_y \sin \theta_4 + {l_5}_x)^2 +
({l_4}_x \sin \theta_4 + {l_4}_y \cos \theta_4 + {l_5}_y)^2 \ .
\eeq
But this distance can also be written as
\bey{A2.7}
q_1^2 &=& \vert {\bf x}- {\bf T}_2 {\bf l}_3 \vert^2 \nonumber \\
{\bf x} &=& {\bf T}_1^{-1} {\bf t} - {\bf l}_2  \ .
\eey
Equating these two results, two values of $\phi_2$ are possible
\bey{A2.8}
\phi_2^{\rm I} &=& 
\arcsin c_1 - \arctan x_y / x_z - H(x_z) \nonumber \\
\phi_2^{\rm II} &=& 
\pi - \arcsin c_1 - \arctan x_y / x_z - H(x_z) \ ,
\eey
with
\beq{A2.9}
H(x) = \left\{
\begin{array}{l}
0,~ x>0 \\[.2in]
\pi,~x<0
\end{array}  \right. \ .
\eeq
The constant $c_1$ is given by
\beq{A2.10}
c_1 = 
\left\{
\begin{array}{l}
\frac{
q_1^2 - x^2 - l_3^2 + 2 x_x
(\cos \theta_2 {l_3}_x + \sin \theta_2 {l_3}_y)
}{-2
 (\sin \theta_2 {l_3}_x - \cos \theta_2 {l_3}_y)
 (x_y^2 + x_z^2)^{1/2}}
,~\theta_3 \ne 0, \theta_5 \ne 0 \\[.2in]
 \frac{{l_3}_x + {l_4}_x + {l_5}_x \cos \theta_4 - x_x \cos \theta_2}
{\sin \theta_2 (x_y^2 + x_z^2)^{1/2}} 
,~\theta_3 = 0, \theta_5 \ne 0 \\[.2in]
\frac{({\bf r}_5 - {\bf r}_2) \cdot ({\bf r}_6 - {\bf r}_5)/l_6
- {l_5}_x - {l_4}_x \cos \theta_4
- x_x (\cos \theta_2 {l_3}_x + \sin \theta_2 {l_3}_y)}
{(\sin \theta_2 {l_3}_x - \cos \theta_2 {l_3}_y)
 (x_y^2 + x_z^2)^{1/2}}
,~\theta_3 \ne 0, \theta_5 = 0 \\[.2in]
\frac{{l_3}_x \cos \theta_4
- x_x (\cos \theta_2 {l_3}_x + \sin \theta_2 {l_3}_y)}
{(\sin \theta_2 {l_3}_x - \cos \theta_2 {l_3}_y)
 (x_y^2 + x_z^2)^{1/2}}
,~\theta_3 = 0, \theta_5 = 0 \\[.2in]
\end{array}  \right. \ ,
\eeq
where ${\bf x}$ is given by Eq.\ (\ref{A2.7}) if $\theta_5 \ne 0$, and
${\bf x} = {\bf T}_1^{-1} [{\bf T}_1^{lab}]^{-1} ({\bf r}_6
- {\bf r}_5) / l_6$ if $\theta_5 = 0$.
Clearly for there to be a solution $\vert c_1 \vert \le 1$.
The last three equations for $c_1$ were determined by conditions
similar to equating Eqs.\ (\ref{A2.6}) and (\ref{A2.7}).
For $\theta_3 = 0, \theta_5 \ne 0$, 
the x-component of
${\bf r}_5^{(3)} - {\bf r}_3^{(3)}$ is known to be equal to
${l_4}_x + {l_5} \cos \theta_4$.
For $\theta_3 \ne 0, \theta_5 = 0$, 
the x-component of
${\bf r}_5^{(5)} - {\bf r}_3^{(5)}$ is known to be equal to
${l_5}_x + {l_4}_x \cos \theta_4$.
For $\theta_3 = 0, \theta_5 = 0$, 
the angle between ${\bf r}_3 - {\bf r}_2$ and
${\bf r}_6 - {\bf r}_5$ is known to be equal to $\theta_4$.

To determine $\phi_3$,  two expressions for $\vert {\bf r}_5 - {\bf
r}_4 \vert ^2$ are again equated to determine
\beq{A2.11}
c_2 = \frac{
l_5^2 - y^2 - l_4^2 + 2 y_x
(\cos \theta_3 {l_4}_x + \sin \theta_3 {l_4}_y)}
{-2
 (\sin \theta_3 {l_4}_x - \cos \theta_3 {l_4}_y)
 (y_y^2 + y_z^2)^{1/2}}
\eeq
and
\bey{A2.12}
\phi_3^{\rm I} &=& 
\arcsin c_2 - \arctan y_y / y_z - H(y_z) \nonumber \\
\phi_3^{\rm II} &=& 
\pi - \arcsin c_2 - \arctan y_y / y_z - H(y_z) \ ,
\eey
where ${\bf y} = {\bf T}_2^{-1} ({\bf T}_1^{-1} {\bf t} - {\bf l}_2)
- {\bf l}_3$.
Again, $\vert c_2 \vert \le 1$ for there to be a solution.

If $\theta_5 \ne 0$,
the value of $\phi_4$ can be determined from
\beq{A2.12a}
{\bf r}_5^{(1)} = 
{\bf r}_4^{(1)}  +  
{\bf T}_1 {\bf T}_2 {\bf T}_3 {\bf T}_4 {\bf l}_5 \ .
\eeq
Defining
\beq{A2.12b}
{\bf q}_3 = 
{\bf T}_3^{-1} {\bf T}_2^{-1} {\bf T}_1^{-1} [{\bf T}_1^{lab}]^{-1}
({\bf r}_5 - {\bf r}_4) \ ,
\eeq
the equations that define $\phi_4$ are given by
\bey{A2.12c}
{q_3}_y &=& \cos \phi_4 (\sin \theta_4 {l_5}_x - \cos \theta_4 {l_5}_y)
\nonumber \\ 
{q_3}_z &=& \sin \phi_4 (\sin \theta_4 {l_5}_x - \cos \theta_4 {l_5}_y)
 \ .
\eey
This is a successful rotation if the position of ${\bf r}_6$ is
successfully predicted.  That is, the equation 
\beq{A2.13}
{\bf r}_6^{(1)} - {\bf r}_5^{(1)} = 
{\bf T}_1 {\bf T}_2 {\bf T}_3 {\bf T}_4{\bf T}_5
{\bf l}_6 = [{\bf T}_1^{lab}]^{-1} ({\bf r}_6 - {\bf r}_5)
\eeq
must be satisfied.  We consider the x-component which implies
\beq{A2.14}
F(\phi_1) = 
\left\{
\begin{array}{l}
({\bf r}_6^{(1)} - {\bf r}_5^{(1)})^\top
{\bf T}_1 {\bf T}_2 {\bf T}_3 {\bf T}_4 \hat {\bf x}
- ({l_6}_x \cos \theta_5 + {l_6}_y \sin \theta_5) = 0
,~ \theta_5 \ne 0 \\[.2in]
({\bf r}_4 - {\bf r}_3) \cdot ({\bf r}_6 - {\bf r}_5) - l_4 l_6 \cos \theta_4
 = 0
,~ \theta_3 \ne 0, \theta_5 = 0 \\[.2in]
\vert {\bf r}_6 - {\bf r}_4\vert - \left[ 
({l_6}_x + {l_5}_x)^2 + {l_5}_y^2\right]^{1/2} = 0
,~ \theta_3 = 0, \theta_5 = 0 \\[.2in]
\end{array}  \right. \ .
\eeq
must be satisfied if the rotation is successful.  The equations for the
case
$\theta_5 = 0$ clearly express the geometric conditions required for
a successful rotation.

Eq.\ (\ref{A2.14}) is the nonlinear
equation for $\phi_1$ that must be solved.  The equation depends only
on $\phi_1$ because $\phi_2$, $\phi_3$, and $\phi_4$ are determined
by Eqs.\ (\ref{A2.8}), (\ref{A2.12}), and (\ref{A2.12c}) in terms of
$\phi_1$.  This equation has between zero and four values for each value
of $\phi_1$, however, due to the multiple root character of Eqs.\
(\ref{A2.8}) and (\ref{A2.12}).  Equation (\ref{A2.14})
is solved by searching
the region $-\pi < \phi < \pi$ for zero crossings.  The search is in
increments of $\approx 0.04^\circ$.  These roots are then refined
by a bisection method.  There is always an even number of roots,
due to the periodic nature of Eq.\ \ref{A2.14}.

The root positions, ${\bf r}_i$, are enough to determine the position
and orientation of the seven rigid units that are modified by the
concerted rotation.  
Rigid unit 0 is translated so that its root position is at ${\bf r}_0$.
It is oriented so that its incoming
bond vector is along the outgoing bond vector of rigid unit $-1$.
It is then rotated so that its outgoing bond vector ends at ${\bf r}_1$.
This process is repeated sequentially for rigid units 1 to 6.

Repeated application of the concerted rotation leads to a slightly
imperfect structure, due to numerical precision errors.  In a practical
application, the geometry would be restored to an ideal state by
application of the SHAKE \cite{Ryckaert}
or Random Tweak algorithm \cite{Shenkin}.

The transformation from $\phi_i,~ 0 \le i \le 6$, to the new solution
which is constrained to change only ${\bf r}_i, ~1 \le i \le 4$,
actually implies a change in volume element in torsional angle space.
This change in volume element is the reason for the appearance of the
Jacobian in the acceptance probability.  The Jacobian of the
transformation for alkanes
is calculated in Ref.\ \cite{Dodd}.  It is slightly
different here since root position ${\bf r}_5$ is not necessarily
the head position.  The Jacobian
is given by
\beq{A2.15}
J = 1/\vert \det {\bf B} \vert \ ,
\eeq
where the $5 \times 5$ matrix $B_{ij}$ is given by the $i$th component of
${\bf u}_j \times ({\bf r}_5 - {\bf h}_j)$ for $i \le 3$ and
by the $(i-3)$th component of 
${\bf u}_j \times ({\bf r}_6 - {\bf r}_5)/
\vert {\bf r}_6 - {\bf r}_5 \vert $ for $i=4, 5$.  Here
${\bf h}_i = {\bf r}_i$ except that ${\bf h}_5$ is the head position
even if $\theta_5 = 0$, and ${\bf u}_i$ is the incoming unit bond
vector for unit $i$.

\section*{Figure Captions}
\flushleft{Figure 1.}
The type I move applied to the serine side
chain.

\flushleft{Figure 2.}
The type IIb move is illustrated for
the case where unit 0 is (a) a B-unit and (b) a C-unit.  In each
case, the original geometry and the four possible new geometries for
the chosen driver angle are shown.  In case (a), one of the new
geometries is very different from the original and the other three
new ones.  The move is shown for a linear peptide, although it is
used only on cyclic peptides.

\flushleft{Figure 3.}
The energy of G$_6$ as a function
of Monte Carlo steps.  Note the rapid equilibration.

\flushleft{Figure 4.}
The probability distribution for the
end-to-end distance for G$_6$.  The distance is between the terminal
C$_\alpha$ groups.

\flushleft{Figure 5.}
The energy of CG$_6$C as a function
of Monte Carlo steps.  Note the rapid equilibration.

\flushleft{Figure 6.}
The number of new solutions found
for each attempted concerted rotation for CG$_6$C.

\flushleft{Figure 7.}
The observed probability distribution  for the
C$_\beta$SSC$_\beta$ torsional angle in CG$_6$C is shown.

\flushleft{Figure 8.}
The potential of mean force calculated by umbrella sampling for 
the C$_\beta$SSC$_\beta$ torsional angle in CG$_6$C (dashed line).
The potential of mean force implied by Fig.\ 7 is indicated by the
solid line.  Also shown is the
bare torsional energy contribution for this rotation (dotted
line).

\flushleft{Figure 9.}
Shown are the occupation numbers of the
configuration in each of the three dominant conformations as
a function of Monte Carlo steps (a).  Also shown is the all-atom
root-mean-square displacement of the configuration from each
of the three dominant conformations (b).  The curves for
conformation 1 are solid, those for 2 are dashed, and those
for 3 are short-dashed.
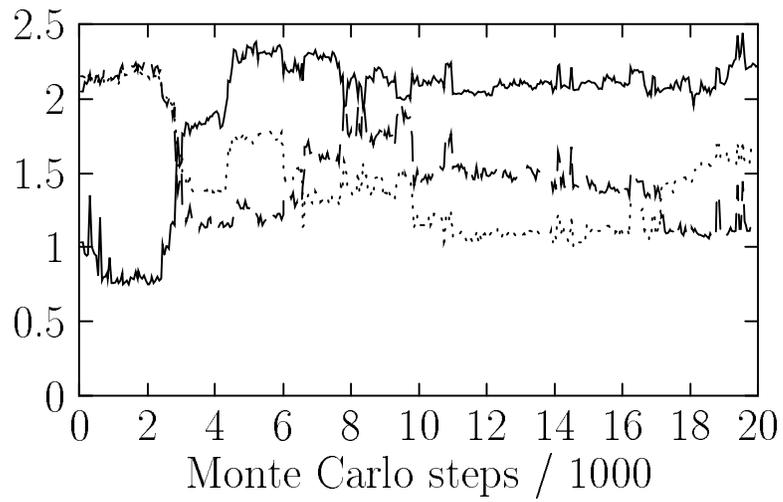
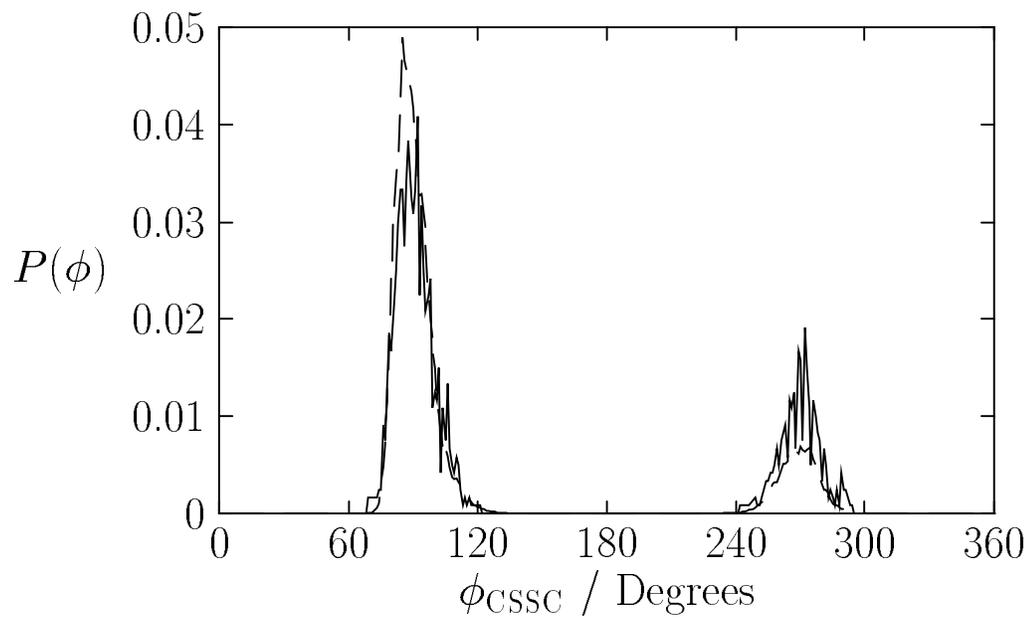
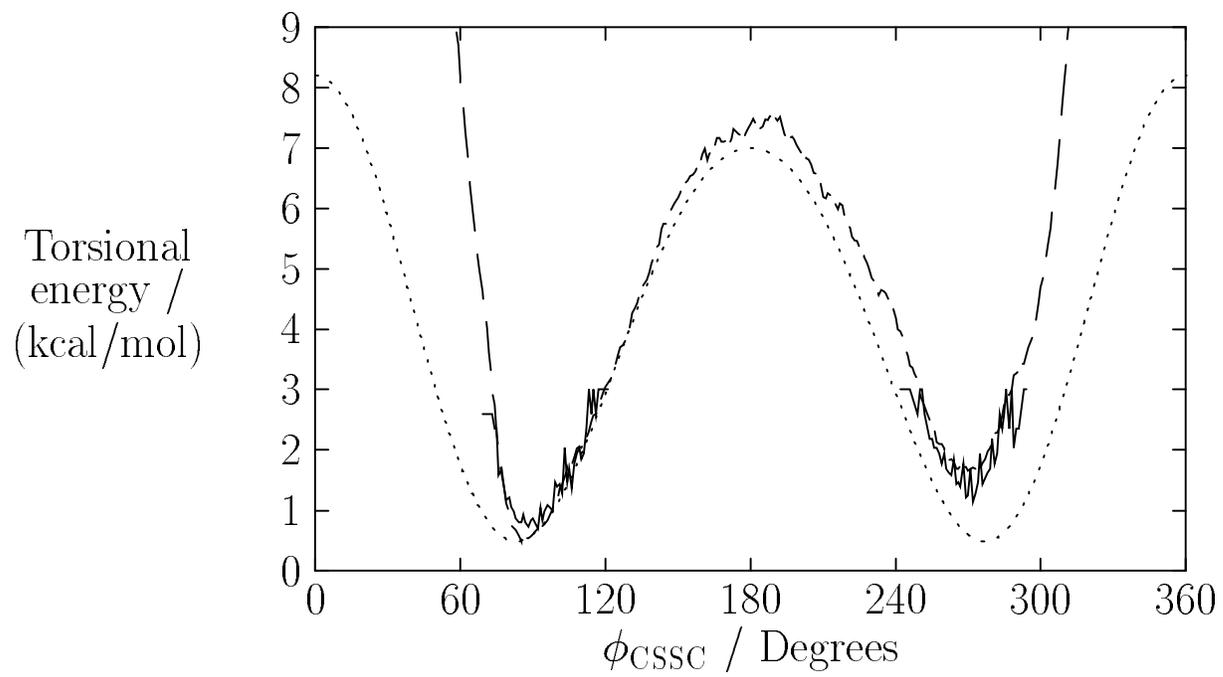
\end{document}